\documentclass[lettersize,journal]{IEEEtran}
\usepackage{amsmath,amsfonts}
\usepackage{algorithmic}
\usepackage{algorithm}
\usepackage{array}
\usepackage[caption=false,font=normalsize,labelfont=sf,textfont=sf]{subfig}
\usepackage{textcomp}
\usepackage{stfloats}
\usepackage{url}
\usepackage{verbatim}
\usepackage{graphicx}
\usepackage{cite}
\usepackage[nolist]{acronym}
\usepackage{siunitx}
\usepackage{booktabs}
\usepackage{multirow}
\usepackage{array}
\usepackage{flushend}
\newlength\mylongcol
\newlength{\mysep}
\IEEEoverridecommandlockouts
\IEEEpubid{%
  \makebox[\textwidth][c]{%
    \begin{minipage}{\textwidth}
      \centering\scriptsize
      \vspace*{-1.5ex}
      \textcopyright~2026 IEEE. Personal use of this material is permitted. Permission from IEEE must be obtained for all other uses, in any current or future media, including reprinting/republishing this material for advertising or promotional purposes, creating new collective works, for resale or redistribution to servers or lists, or reuse of any copyrighted component of this work in other works.
    \end{minipage}%
  }%
}

\begin{document}

\begin{acronym}[Bash]
\acro{API}{application programming interface}
\acro{AC}{air conditioners}
\acro{DLC}{direct load control}
\acro{DR}{demand response}
\acro{DSO}{distribution system operator}
\acro{DERMS}{Distributed Energy Resource Management System}
\acro{EKZ}{Utility of the Canton of Zurich}
\acro{EV}{electric vehicle}
\acro{EWH}{electric water heater}
\acro{FW}{firmware}
\acro{HEMS}{home energy management system}
\acro{HP}{heat pump}
\acro{KPI}{key performance indicator}
\acro{MIP}{mixed-integer programming}
\acro{MSD}{most similar day}
\acro{PI}{Proportional-Integral}
\acro{PV}{photovoltaic}
\acro{RL}{reinforcement learning}
\acro{SFH}{single-family home}
\acro{SM}{smart meter}
\acro{SoC}{state-of-charge}
\acro{ToU}{Time-of-Use}
\acro{TS}{transformer station}
\acro{TCL}{thermostatically controlled load}
\acro{VAT}{value added tax}
\end{acronym}

\title{Residential Peak Load Reduction via \\ Direct Load Control under Limited Information}

\author{Katharina Kaiser,~\IEEEmembership{Member,~IEEE}, Gustavo Valverde,~\IEEEmembership{Senior Member,~IEEE}, Gabriela Hug,~\IEEEmembership{Senior~Member,~IEEE}%
\thanks{This work is part of a project that received funding from the Swiss Federal Office of Energy~(SFOE) under grant number SI/502271.}%
\thanks{Katharina Kaiser, Gustavo Valverde, and Gabriela Hug are with the Department of Information Technology and Electrical Engineering, ETH Zurich, 8092 Zurich, Switzerland (e-mail: kkaiser@ethz.ch; gustavov@ethz.ch; ghug@ethz.ch).}}

\maketitle
\enlargethispage{-0.3in}

\begin{abstract}
Thermostatically controlled loads and electric vehicles offer flexibility to reduce power peaks in low-voltage distribution networks. This flexibility can be maximized if the devices are coordinated centrally, given some level of information about the controlled devices. In this paper, we propose novel optimization-based control schemes with prediction capabilities that utilize limited information from heat pumps, electric water heaters, and electric vehicles. The objective is to flatten the total load curve seen by the distribution transformer by restricting the times at which the available flexible loads are allowed to operate, subject to the flexibility constraints of the loads to preserve customers' comfort. The original scheme was tested in a real-world setup, considering both summer and winter days.
The pilot results confirmed the technical feasibility but also informed the design of an improved version of the controller. Computer simulations using the adjusted controller show that, compared to the original formulation, the improved scheme achieves greater peak reductions in summer. Additionally, comparisons were made with an ideal controller, which assumes perfect knowledge of the inflexible load profile, the models of the controlled devices, the hot water and space heating demand, and future electric vehicle charging sessions. The proposed scheme with limited information achieves \SI{43.3}{\%} and \SI{27.8}{\%} of the potential average daily peak reduction that the ideal controller with perfect knowledge would achieve in summer and winter, respectively.
\end{abstract}


\section{Introduction}
\label{sec:intro}
\IEEEPARstart{T}{he} electrification of transport and heating in the residential sector, through the adoption of \acp{EV} and \acp{HP}, is a key driver for the energy transition~\cite{OREILLY2024}. This electrification poses significant challenges to the operation of distribution grids, including increased load peaks, temporary equipment overloading, and potential voltage issues~\cite{Damianakis2025}. At the same time, the rapid adoption of rooftop \ac{PV} systems causes new injection peaks, leading to reverse power flows, overloadings, and voltage rises~\cite{Procopiou21}. 

As the heating and transportation sectors exhibit considerable flexibility \cite{IEA2022DER}, demand-side management in the form of direct control of flexible loads has emerged as a promising strategy to address these challenges~\cite{Callaway11}. Heating systems can utilize thermal inertia to shift demand~\cite{Buechler2024}, and \acp{EV} offer great flexibility during the charging process~\cite{Cui2016}. 
\Ac{DLC} can help avoid power peaks in the low-voltage grid and ease the integration of \ac{PV} systems. Ripple control has been used to disconnect \acp{TCL} at specific hours through frequency-ripple signals sent via power cables~\cite{Ross48}. However, demand synchronization after the ripple period may result in undesired load rebounds, highlighting the need for new schemes that can distribute demand more evenly throughout the day. Reference \cite{ozkop2024} provides a comprehensive survey of \ac{DLC} schemes, organized by control objectives, constraints, methods, and appliances -- mainly laundry appliances, \acp{TCL}, and \acp{EV}.
We focus on the latter two for load flattening and peak reductions.
\enlargethispage{-0.6in}

Previous works have proposed direct control of \acp{HP} for temporary load reduction~\cite{GEORGES2017,Ruelens2019}, the control of \acp{EV} for load flattening~\cite{Reddy2020}, the use of \ac{AC} loads to compensate fluctuations in \ac{PV} power~\cite{Mahdavi17}, and load scheduling~\cite{Kun2004, Luo2016}. However, these algorithms require model training or information that is difficult to monitor, collect, and transmit to a central processor. In \cite{SRITHAPON2023}, the authors proposed predictive control of \acp{PV}, \acp{EV}, and \acp{HP} in the same household to reduce electricity bills, instead of part of a central \ac{DLC}. Control schemes with prediction capabilities were also reported in \cite{Starke2020} and \cite{Buechler2025} for load shifting of \acp{EWH}, but they rely on information about devices' temperatures or water consumption.

An optimization-based \ac{DLC} with limited information was proposed in~\cite{Xiang2020} for peak shaving using \acp{EWH}. Water temperatures were estimated using hot-water consumption patterns and tank characteristics. However, the estimation errors accumulate over time, which may affect controller performance. 
 
The work in~\cite{Tulabing23} tested a ripple-like \ac{DLC} scheme to switch off \acp{TCL} and \acp{EV} according to specific thresholds. The controller was validated in a real-world setup involving five houses and then tested in a simulation environment with 70 houses. However, the controller lacked prediction capabilities.

A few pilot projects have evaluated the response of \acp{TCL} to \ac{DLC} commands in real-world conditions \cite{Sullivan13,Desrochers19,Mueller2019,GUPTA2023}. These commands were not optimized to flatten the load curve, but to reduce the demand during specific intervals for potential ancillary services.
Other field tests have been conducted under controlled environments and simulated water consumption patterns of \acp{EWH} \cite{Adham2022}, \cite{Zeng2024}. These pilots did not assess peak demand reductions, but focused on the implications for the enrolled customers and information-exchange testing.  

Previous \acp{DLC} involve mainly homogeneous devices, but the coordination of  \acp{HP} with \acp{EWH} and \acp{EV} is expected to maximize their flexibility  \cite{Aschidamini2025}. Although a limited number of \acp{DLC} possess prediction capabilities, they fail to explicitly account for past commands in their current and future decision-making processes. Additionally, most \acp{DLC} require measurements or estimations of water and room temperatures, which may be unavailable or difficult to obtain in real implementations~\cite{Cui2025}. Therefore, new \acp{DLC} must be tailored to more realistic conditions. Finally, the majority of the proposed formulations lack verification in real-world conditions, hiding potential implementation issues and depriving the formulation of improvements where needed.
 
This paper proposes two \acp{DLC} designed to reduce \ac{PV} injection and demand peaks in low-voltage networks. The schemes optimize the blocking of \acp{HP}, \acp{EWH}, and \acp{EV} in a 24-hour rolling horizon, requiring only limited knowledge of their operating conditions, reflected in flexibility constraints. This approach avoids the need for temperature sensors or detailed \ac{TCL} thermal models. Moreover, the proposed schemes take into account past commands and predictions of the inflexible load over the next 24 hours. The first version of the controller was implemented in a real-world setup, specifically in a pilot project that controlled tens of devices in a small community in the canton of Zurich. This pilot was part of OrtsNetz, a project carried out by ETH Zurich in collaboration with the \ac{EKZ}, which partnered on the pilot implementation and coordination with participating customers, and provided valuable inputs to keep the \acp{DLC} practical, minimally disruptive to customers' comfort, and fully implementable in real life.
Interested readers may consult our technical report \cite{KaiserOrtsNetz2025}, the final deliverable for the Swiss Federal Office of Energy, for further information about the project, covering customer participation, the design and implementation of indirect schemes based on real-time and time-of-use tariffs, as well as preliminary versions of the \acp{DLC} refined in this paper. 

The pilot results showed technical feasibility, but also revealed the need for a refined optimization formulation. The improved scheme addresses the identified drawbacks and was tested in a simulation environment similar to that of the pilot project. Hence, the contributions of this paper are:
\begin{itemize}
    \item Derivation of new centralized control schemes (L1) and (L2) of residential \acp{HP}, \acp{EWH}, and \acp{EV} to reduce peak loads using limited information. These formulations take into account the flexibility constraints of the controlled devices to guarantee end-user comfort. 
    \item Implementation and performance evaluation of  formulation (L1) in a pilot project with 33 \acp{TCL} from 22 houses and up to two simultaneous \ac{EV} charging sessions. Incorporation of the lessons learned from the real-world setup in the refined formulation (L2). 
    \item Impact assessment of the controllers' performance for different information availability conditions. The performance is compared to an ideal setup with perfect knowledge.
\end{itemize}

The \acp{DLC} presented in this paper resemble unit commitment problems with minimum/maximum up/down times of flexible loads. However, they only block the \ac{TCL} operation, while the internal thermostat handles the transition to the on state. These formulations
were designed with the pilot project’s communication, information, and
actuation capabilities in mind, making them readily available for real-world implementation. These aspects further distinguish our work from that in the literature.

The remainder of this paper is organized as follows: Section~\ref{sec:form} presents the proposed \ac{DLC} formulations with perfect and limited information. Section \ref{sec:res} describes the setup of the pilot project and the performance of (L1) in the real-world implementation. It also presents a comparison of the different \ac{DLC} formulations in a simulation environment with conditions similar to those in the real-world setup. The results are further discussed in Section~\ref{sec:discussion}, and concluding remarks are presented in Section~\ref{sec:conc}.

\section{DLC formulations}
\label{sec:form}
This section presents \ac{DLC} formulations with perfect knowledge and with limited information. The proposed \acp{DLC} are centralized, optimization-based schemes that aim to reduce power peaks at the \ac{TS}. Grid voltages and currents were not monitored or incorporated into the formulations, making them applicable not just for utilities but also aggregators without access to the grid model. For compatibility with different \ac{HP} and \ac{EWH} models, the \acp{DLC} can only block, but not start, device operation, while \ac{EV} charging can be started and stopped.

The controller uses a rolling horizon. At each iteration, it considers time steps $\mathcal{T} = \{0, \dots, T-1\}$ and computes the switching commands for the flexible loads of each customer $c$, but only the first $\kappa$ commands are applied. Device constraints apply to all customers with a given device, though the expression $\forall c \in \cdot$ is omitted in the formulation for simplicity. Power and energy values are in \si{kW} and \si{kWh}, respectively, and each time step has a duration of $\Delta t$ hours.

\subsection{DLC with perfect knowledge}
\label{sec:formIdeal}
This subsection introduces the formulation for the ideal \ac{DLC} which fulfills the same flexibility constraints as the \ac{DLC} with limited information but assumes perfect knowledge of the \ac{HP} and \ac{EWH} models, connection times and charging needs of future \ac{EV} charging sessions, and the level of inflexible load. This formulation serves as a benchmark for the proposed \acp{DLC} with limited information.

\subsubsection{HP constraints}\label{sec:formIdealHP}
The flexibility constraints are derived from the standard time-based ripple control scheme to ensure customer comfort. In this scheme, \acp{HP} can be blocked for up to $K^\text{block,24h}_{c,\text{HP}}$ time steps per 24-hour window:
\begin{align}
    & \sum_{i=t-K+1}^t u_{c,i}^\text{HP}  \geq K-K^\text{block,24h}_{c,\text{HP}}, &&\forall t \in \mathcal{T}
    \label{eq:blockHP}
\end{align}
where $K$ is the number of time steps in 24 hours ($24/\Delta t$). The binary variables $u_{c,t}^\text{HP}$ denote switching commands. For $t<0$, these are inputs to the optimization. When $u_{c,t}^\text{HP}$ is $0$, the \ac{HP} is blocked and cannot operate. When it is $1$, the device may operate or not, depending on the internal device controller.
Additionally, \acp{HP} can only be blocked for up to $K^\text{max,b}_{c,\text{HP}}$ consecutive time steps, and there must be at least $K^\text{min,u}_{c,\text{HP}}$ time steps between two blocking instances:
\newcommand{\hsep}{\hspace{1cm}}
\begin{align}
    \sum_{i=t}^{t+K^\text{max,b}_{c,\text{HP}}} \kern-0.7em u_{c,i}^\text{HP} & \geq w_{c,t}^\text{HP}, \quad \forall t \in \left\{-K^\text{max,b}_{c,\text{HP}}, \dots, T-1-K^\text{max,b}_{c,\text{HP}}\right\} \raisetag{1.5ex}\label{eq:blockinstHP}\\
    \sum_{i=t-K^\text{min,u}_{c,\text{HP}}+1}^{t} \kern-1.3em v_{c,i}^\text{HP} & \leq u_{c,t}^\text{HP}, \quad \forall t \in \mathcal{T} \label{eq:minunblockHP}\\
    u_{c,t}^\text{HP} - u_{c,t-1}^\text{HP} & = v_{c,t}^\text{HP} - w_{c,t}^\text{HP}, \qquad \forall t \in \mathcal{T}\label{eq:switchingHP}
\end{align}
The binary variables $v_{c,t}^\text{HP}$ and $w_{c,t}^\text{HP}$ indicate a change in the blocking state and take the value $1$ at the start of an unblocked and blocked period, respectively. Introducing these additional variables for formulating minimum/maximum up/down times is preferred to formulating the constraints in the $u$-space only, even though it increases the number of binary variables \cite{KnuevenMixedIntegerProgramming2020}.
Finally, a minimum blocking duration of $K^\text{min,b}_{c,\text{HP}}$ time steps is specified:
\begin{align}
    & \sum_{i=t-K^\text{min,b}_{c,\text{HP}}+1}^{t} w_{c,i}^\text{HP} \leq 1 - u_{c,t}^\text{HP}, &&\forall t \in \mathcal{T}\label{eq:minblockHP1}\\
    & \sum_{i=t-K+K^\text{min,b}_{c,\text{HP}}}^{t-1} u_{c,i}^\text{HP} \geq \left(K-K^\text{block,24h}_{c,\text{HP}}\right) \cdot w_{c,t}^\text{HP}, &&\forall t \in \mathcal{T}\label{eq:minblockHP3}
\end{align}

Besides the flexibility constraints \eqref{eq:blockHP} -- \eqref{eq:minblockHP3}, the \ac{HP} constraints in the ideal case also include the underlying \ac{HP} models used in the simulation environment.
Both on-off \acp{HP} and modulating \acp{HP} are modeled with a thermal storage tank. The temperature evolution in the water tank is described by:
\begin{align}
& T_{c,t}^\text{HP}=T_{c,t-1}^\text{HP} + \frac{P_{c,t}^\text{HP} \eta_{c,t}^\text{HP} \Delta t-E_{c,t}^\text{HP}-E_{c,t}^\text{HP,loss}}{V_{c}^\text{HP} \rho_\text{w} c_\text{w} \cdot \frac{1}{3.6 \cdot 10^{6}}}, \; \forall t \in \mathcal{T} \label{eq:tempEvoHP}
\end{align}
where $T_{c,t}^\text{HP}$ is the storage tank temperature at the end of time step~$t$. The first term in the numerator is the thermal energy supplied by the \ac{HP}, where $P_{c,t}^\text{HP}$ is the electrical \ac{HP} power, and $\eta_{c,t}^\text{HP}$ is the \ac{HP}'s coefficient of performance. The parameter $E_{c,t}^\text{HP}$ is the thermal energy withdrawn for space heating, and $E_{c,t}^\text{HP,loss}=\lambda_c^\text{HP}\left(T_{c,t-1}^\text{HP}-T_c^\text{env}\right)\Delta t$ are the heat losses, where $\lambda_c^\text{HP}$ is the thermal conductance in \si{kW.K^{-1}} and $T_c^\text{env}$ is the room temperature around the tank.
$V_{c}^\text{HP}$ is the tank size in \si{m^3}, $\rho_\text{w}=\SI{1000}{kg.m^{-3}}$ is the water density, $c_\text{w}=\SI{4184}{J.kg^{-1}.K^{-1}}$ is the specific heat capacity, and the fraction $1/(3.6 \cdot 10^{6})$ converts \si{Ws} to \si{kWh}.

The \ac{HP} power is subject to the operating state $\rho_{c,t}^\text{HP}$, which depends on the internal hysteresis controller and the external blocking signal:
\begin{align}
& \rho_{c,t}^\text{HP}-\rho_{c, t-1}^\text{HP} \leq \frac{T_{c,t}^\text{HP,LO}\!\!-T_{c,t-1}^\text{HP}}{M_1} + 1,
    &&\forall t \in \mathcal{T} \label{eq:hystHP1} \\
& \rho_{c,t}^\text{HP}-\rho_{c, t-1}^\text{HP} \geq\frac{T_{c,t}^\text{HP,UP}\!\!-T_{c,t-1}^\text{HP}}{M_2} - 1  \nonumber\\
    & \kern9.3em + u_{c,t}^\text{HP}-1, &&\forall t \in \mathcal{T} \label{eq:hystHP2} \\
& \rho_{c,t}^\text{HP} \geq \frac{T_{c,t}^\text{HP,LO}\!\!-T_{c, t-1}^\text{HP}}{M_3} + u_{c,t}^\text{HP}-1,
    &&\forall t \in \mathcal{T} \label{eq:hystHP3} \\
& \rho_{c,t}^\text{HP} \leq \frac{T_{c,t}^\text{HP,UP}\!\!-T_{c, t-1}^\text{HP}}{M_4} + 1,
    &&\forall t \in \mathcal{T} \label{eq:hystHP4}\\
& \rho_{c,t}^\text{HP} \leq u_{c,t}^\text{HP},
    &&\forall t \in \mathcal{T} \label{eq:RhoSmallerHeatsysactUHPideal}
\end{align}
Constraints \eqref{eq:hystHP1} -- \eqref{eq:hystHP4} without the term $u_{c,t}^\text{HP}-1$ in \eqref{eq:hystHP2} and \eqref{eq:hystHP3} model the hysteresis control. Constraints \eqref{eq:hystHP1} and \eqref{eq:hystHP2} keep the operating state unchanged when the temperature is between the lower and upper temperature limits ($T_{c,t}^\text{HP,LO}$ and $T_{c,t}^\text{HP,UP}$), while \eqref{eq:hystHP3} and \eqref{eq:hystHP4} turn it on (resp. off) when the temperature is below (resp. above) the hysteresis band. The constants $M_1$ -- $M_4$ must be greater than the expected maximum temperature differences, such that the absolute values of the fractions in \eqref{eq:hystHP1} -- \eqref{eq:hystHP4} are $\leq 1$.
The \ac{DLC} can overrule the \ac{HP} hysteresis control by blocking device operation. When $u_{c,t}^\text{HP}=0$, constraint \eqref{eq:RhoSmallerHeatsysactUHPideal} forces the \ac{HP} to be off independently of the temperature. To ensure the feasibility of the other constraints during a blocking event, the term $u_{c,t}^\text{HP}-1$ in \eqref{eq:hystHP2} and \eqref{eq:hystHP3} enables turning off even though the temperature is within the specified limits and being off even though the temperature is below the lower limit. After a blocking event, the \ac{HP} remains off until the temperature drops below the lower bound.

For on-off \acp{HP}, the power in time step $t$ is assumed to be either $0$ or the nominal $P_{c,\text{nom}}^\text{HP}$, depending on the binary variable $\rho_{c,t}^\text{HP}$, i.e.,
\begin{align}
    P_{c,t}^\text{HP} &= \rho_{c,t}^\text{HP} \cdot P_{c,\text{nom}}^\text{HP}, &&\forall t \in \mathcal{T} \label{eq:POnOffHPideal}
\end{align}
Modulating \acp{HP} are assumed to match the sum of the thermal demand, the losses, and the energy required to reach the lower temperature bound, described by auxiliary variables $E_{c,t}^\text{HP,LO}$. The resulting power $\hat{P}_{c,t}^\text{HP}$ then needs to be mapped to the feasible power range $[P_{c,\text{min}}^\text{HP}, P_{c,\text{nom}}^\text{HP}]$ of the \ac{HP}. The following constraints reflect this behavior:
\begin{align}
    &E_{c,t}^\text{HP,LO}\!\!=\!\max\!\left(\!0,\!\left(T_{c,t}^\text{HP,LO}\!\!-T_{c, t-1}^\text{HP}\right)\!\cdot\!\frac{V_{c}^\text{HP} \rho_\text{w} c_\text{w}}{3.6 \cdot 10^{6}}\right)\!, \hspace{-0.25cm}&&\forall t\!\in\!\mathcal{T}\label{eq:EauxModHPideal}\\
    &\hat{P}_{c,t}^\text{HP} = \frac{E_{c,t}^\text{HP} + E_{c,t}^\text{HP,loss} + E_{c,t}^\text{HP,LO}}{\eta_{c,t}^\text{HP} \cdot \Delta t}, &&\forall t\!\in\!\mathcal{T}\label{eq:PauxModHPideal}\\
    &P_{c,t}^\text{HP} = \rho_{c,t}^\text{HP} \cdot 
        \max \left( 
            P_{c,\text{min}}^\text{HP}, 
            \min \left( P_{c,\text{nom}}^\text{HP}, \hat{P}_{c,t}^\text{HP} \right)
        \right), \hspace{-0.25cm}&&\forall t\!\in\!\mathcal{T}\label{eq:PModHPideal}
\end{align}

\subsubsection{EWH constraints}
Similar to \acp{HP}, \acp{EWH} can be blocked for up to $K^\text{block,24h}_{c,\text{EWH}}$ time steps per 24-hour window:
\begin{align}
    & \sum_{i=t-K+1}^t u_{c,i}^\text{EWH}  \geq K-K^\text{block,24h}_{c,\text{EWH}}, &&\forall t \in \mathcal{T}
    \label{eq:blockWH}
\end{align}

The \ac{EWH} model in the ideal \ac{DLC} is the same as the model for on-off \acp{HP}, except that (i)~the coefficient of performance is assumed to be $1$ and is therefore discarded, and (ii)~the temperature limits are time-independent for \acp{EWH}. Hence, the constraints include \eqref{eq:blockWH} and \eqref{eq:tempEvoHP} -- \eqref{eq:POnOffHPideal} adapted to \acp{EWH}.

\subsubsection{EV constraints}
The ideal \ac{DLC} takes into account all controllable charging sessions $\mathcal{S}_c = \{0, \dots, S_c-1\}$ of each customer $c$ in the optimization horizon.
The first time step of each session~$s$ is denoted as $t_{c,s}^\text{EV,a}$. It corresponds to the start time of a future charging session or time step $0$ if the session is ongoing. Additionally, the customer specifies the departure time step $t_{c,s}^\text{EV,d}$ and the corresponding desired \acl{SoC} $\text{SoC}_{c,s}^\text{EV,goal}$ for each controllable charging session. Given the \acs{SoC} at the end of time step $t_{c,s}^\text{EV,a} - 1$, the desired \acs{SoC}, the charging efficiency $\eta_c^\text{EV}$, the nominal charging power $P_{c,\text{nom}}^\text{EV}$, and the battery capacity $E_{c}^\text{EV}$, the number of time steps for which the \ac{EV} needs to charge until departure is:
\begin{align}
    \kern-0.5em N_{c,s}^\text{EV,goal} \! =  
    \!\! \left\lceil \!
    \left(\text{SoC}_{c,s}^\text{EV,goal} - \text{SoC}_{c,t_{c,s}^\text{EV,a}-1}^\text{EV}\right) \kern-0.1em 
    \cdot \kern-0.1em \frac{E_{c}^\text{EV}}
    {P_{c,\text{nom}}^\text{EV} \kern-0.1em \cdot \kern-0.1em \eta_{c}^\text{EV} \kern-0.1em \cdot \kern-0.1em \Delta t}
    \! \right\rceil
\end{align}
and the number of charging time steps to reach a given minimum \acs{SoC} is:
\begin{align}
    \kern-0.5em \hat{N}_{c,s}^\text{EV,min} \! =  
    \!\! \left\lceil \!
    \left(\text{SoC}_c^\text{EV,min} - \text{SoC}_{c,t_{c,s}^\text{EV,a}-1}^\text{EV}\right) \kern-0.1em 
    \cdot \kern-0.1em \frac{E_{c}^\text{EV}}
    {P_{c,\text{nom}}^\text{EV} \kern-0.1em \cdot \kern-0.1em \eta_{c}^\text{EV} \kern-0.1em \cdot \kern-0.1em \Delta t}
    \! \right\rceil
\end{align}
In case the desired \acs{SoC} is smaller than the minimum \acs{SoC}, charging should be stopped after reaching the desired \acs{SoC}, with $N_{c,s}^\text{EV,min}\!=\!\min(\hat{N}_{c,s}^\text{EV,min}\!\!, N_{c,s}^\text{EV,goal})$. These values are used in the following constraints to enforce that (i)~the \ac{EV} reaches the desired \acs{SoC} at departure, and (ii)~charging is only stopped after reaching the minimum \acs{SoC}, which is introduced to reduce customers' dissatisfaction in case of communication issues:
\renewcommand{\hsep}{\hspace{2cm}}
\begin{align}
    &\sum_{i=t_{c,s}^\text{EV,a}}^{t_{c,s}^\text{EV,d}-1} u_{c,i}^\text{EV} = N_{c,s}^\text{EV,goal}, \quad \forall s\in \mathcal{S}_c \text{ with } t_{c,s}^\text{EV,d} \leq T \label{eq:SoCgoalEV1}\\
    &\sum_{i=t_{c,s}^\text{EV,a}}^{T-1} u_{c,i}^\text{EV} \leq N_{c,s}^\text{EV,goal}, \quad \forall s\in \mathcal{S}_c \text{ with } t_{c,s}^\text{EV,d} > T \label{eq:SoCgoalEV2}\\
    &\sum_{i=t_{c,s}^\text{EV,a}}^{t_{c,s}^\text{EV,a} + N_{c,s}^\text{EV,min}-1} \left(1-u_{c,i}^\text{EV}\right) = 0, \nonumber\\[-1.5ex]
        &\hsep\forall s\in \mathcal{S}_c \text{ with } t_{c,s}^\text{EV,a} + N_{c,s}^\text{EV,min} \leq T \label{eq:SoCminEV1}\\
    &\sum_{i=t_{c,s}^\text{EV,a}}^{T-1} \left(1-u_{c,i}^\text{EV}\right) = 0, \nonumber\\[-1.5ex]
        &\hsep\forall s\in \mathcal{S}_c \text{ with } t_{c,s}^\text{EV,a} + N_{c,s}^\text{EV,min} > T \label{eq:SoCminEV2}\\
    &u_{c,t}^\text{EV} \leq \phi_{c,t}^\text{EV}, \hspace{0.3cm} \forall t\in \mathcal{T} \label{eq:EVhome}
\end{align}
The binary parameter $\phi_{c,t}^\text{EV}$ in \eqref{eq:EVhome} denotes whether the \ac{EV} is plugged in, and the binary variable $u_{c,t}^\text{EV}$ denotes whether it charges.
Additionally, communication capabilities limit the number of off-to-on switchings in any 24-hour window to $K^\text{max,24h}_{c,\text{EV}}$:
\begin{align}
    & \sum_{i=t-K+1}^t b_{c,i}^\text{EV} \leq K^\text{max,24h}_{c,\text{EV}},\quad &&\forall t \in \mathcal{T}
    \label{eq:maxEVswitch}\\
    & b_{c,t}^\text{EV} \geq \left(u_{c,t}^\text{EV} - u_{c,t-1}^\text{EV}\right) \cdot \phi_{c,t-1}^\text{EV}, &&\forall t \in \mathcal{T}
    \label{eq:auxvarEV}
\end{align}
where the binary variable $b_{c,t}^\text{EV}$ describes an off-to-on transition. Upon arrival, the \ac{EV} starts charging by default, which is not counted as a start signal. Finally, each \ac{EV} is assumed to charge with the given nominal power when $u_{c,t}^\text{EV} = 1$:
\begin{align}
    P^\text{EV}_{c,t} &= u_{c,t}^\text{EV} \cdot P_{c,\text{nom}}^\text{EV}, &&\forall t \in \mathcal{T}\label{eq:powerEV}
\end{align}
Potential average power deviations in the last time step due to the battery being full are neglected.

\subsubsection{Objective function}
The ideal \ac{DLC} minimizes the highest absolute total power in the optimization horizon:
\begin{subequations}
\begin{align}
\min \;\;  &P^\text{max}\\
\text{s.t.} \;\: \,
& P^\text{max} \!\geq P_t^\text{tot} 
    , \kern3em \forall t \in \mathcal{T}\label{eq:ideal_obj_aux1}\\
& P^\text{max} \!\geq - P_t^\text{tot}
    , \kern2.25em \forall t \in \mathcal{T}\label{eq:ideal_obj_aux2}\\
& P_t^\text{tot} \!= P_{t}^\text{inf}
    + \!\!\sum_{c\in \mathcal{C^\text{HP}}}\!\! P^\text{HP}_{c,t} 
    + \!\!\sum_{c\in \mathcal{C^\text{EWH}}}\!\! P^\text{EWH}_{c,t}
    + \!\!\sum_{c\in \mathcal{C^\text{EV}}}\!\! P^\text{EV}_{c,t} 
    , \, \forall t \in \mathcal{T}\raisetag{1.5ex}\label{eq:ideal_obj_ptot}
\end{align}
\label{eq:ideal_obj}
\end{subequations}

\noindent where $P^\text{inf}_t$ is the inflexible load at \ac{TS} level, i.e., the load that is not controlled by the \ac{DLC}, and $\mathcal{C^\text{HP}}$, $\mathcal{C^\text{EWH}}$, and $\mathcal{C^\text{EV}}$ are the sets of customers with a corresponding controllable device.

\subsection{DLC with limited information}
\label{sec:formLimited}
In the given real-world setup, the central controller knows the nominal power values for \acp{HP} and \acp{EWH}, as well as the \ac{EV} charging demand during ongoing charging sessions. However, it lacks knowledge of future charging sessions and \ac{TCL} temperatures, and relies on estimates for the inflexible power $P^\text{inf}_t$ at the \ac{TS} level, as well as the \ac{HP} and \ac{EWH} demand. Hence, the formulation of the optimization problem needs to be adapted.

\subsubsection{HP constraints}
The \ac{HP} constraints include \eqref{eq:blockHP} -- \eqref{eq:minblockHP3} for all customers with an \ac{HP}, and the power is estimated based on the ambient temperature, represented by the parameter $0 \leq \alpha_t^\text{HP} \leq 1$ (cf.~Section~\ref{sec:HP_power_estimation}):
\begin{align}
    P^\text{HP}_{c,t} &= u_{c,t}^\text{HP} \cdot \alpha_t^\text{HP} \cdot P_{c,\text{nom}}^\text{HP},
    &&\forall t \in \mathcal{T} \label{eq:powerHP}
\end{align}

\subsubsection{EWH constraints}
Two different formulations to model \ac{EWH} operation are investigated. The optimization problems corresponding to these two formulations are referred to as (L1) and (L2). Constraint \eqref{eq:blockWH} applies in both cases.

In (L1), constraints \eqref{eq:blockinstHP} -- \eqref{eq:minblockHP3}, which specify a minimum and maximum blocking duration and a minimum duration between two blocking instances, are also introduced for \acp{EWH}. The inclusion of a minimum blocking duration reduces the uncertainty related to \acp{EWH} being on when unblocked because after a long blocking period, implemented by choosing a high value for $K^\text{min,b}_{c,\text{EWH}}$, an \ac{EWH} is likely to operate as soon as it is unblocked. The \ac{EWH} power demand is:
\begin{align}
    P^\text{EWH}_{c,t} &= u_{c,t}^\text{EWH} \cdot P_{c,\text{nom}}^\text{EWH},
    &&\forall t \in \mathcal{T} \label{eq:powerEWH1}
\end{align}
This formulation was tested in the pilot project, as described in Section \ref{sec:resPil}.
An alternative, improved formulation (L2) is tested in Section~\ref{sec:resSim}. It includes additional binary variables $\rho_{c,t}^\text{EWH}$ that model the \ac{EWH} power demand when the device is unblocked. 
We assume that the \ac{EWH} runs during the first $K^\text{run}_{c,\text{EWH}}$ time steps of an unblocking period and demands the nominal power $P_{c,\text{nom}}^\text{EWH}$. 
The following constraints enforce $\rho_{c,t}^\text{EWH}=1$ for the first $K^\text{run}_{c,\text{EWH}}$ time steps of an unblocking period, and $0$ otherwise:\footnote{An alternative formulation is $\rho_{c,t}^\text{EWH} = \sum_{i=t-K^\text{run}_{c,\text{EWH}}+1}^{t} v_{c, i}^\text{EWH}$; however, with the studied settings, the given formulation reduced computation time.}
\renewcommand{\hsep}{\hspace{1cm}}
\begin{align}
\rho_{c,t}^\text{EWH} &\geq \sum_{i=t-K^\text{run}_{c,\text{EWH}}+1}^{t} v_{c, i}^\text{EWH}, &&\forall t \in \mathcal{T} \label{eq:Krunrho1EWH} \\
\rho_{c,t}^\text{EWH} &\leq 1 - \sum_{i=t-K^\text{max,u}_{c,\text{EWH}}+1}^{t-K^\text{run}_{c,\text{EWH}}} v_{c, i}^\text{EWH}, &&\forall t \in \mathcal{T} \label{eq:Krunrho0EWH} \\
\rho_{c,t}^\text{EWH} &\leq u_{c,t}^\text{EWH}, &&\forall t \in \mathcal{T} \label{eq:RhoSmallerUEWH}
\end{align}
To model that the \ac{EWH} may still operate during some of the remaining unblocked time steps, we account for \SI{10}{\%} of the nominal power when $u_{c,t}^\text{EWH}=1$ and $\rho_{c,t}^\text{EWH}=0$:
\begin{align}
    P^\text{EWH}_{c,t} &= \left(0.1 \cdot u_{c,t}^\text{EWH} + 0.9 \cdot \rho_{c,t}^\text{EWH}\right) \cdot P_{c,\text{nom}}^\text{EWH}, \; \forall t \in \mathcal{T} \label{eq:powerEWH2}
\end{align}
In contrast to (L1), the constraint set includes only \eqref{eq:minunblockHP} -- \eqref{eq:minblockHP3} adapted to \acp{EWH}. Furthermore, an upper limit on the number of consecutive unblocked time steps is added:
\begin{align}
    & \sum_{i=t}^{t+K^\text{max,u}_{c,\text{EWH}}} \left(1 - u_{c,i}^\text{EWH}\right) \geq v_{c,t}^\text{EWH}, \nonumber\\
        & \hsep \forall t \in \left\{-K^\text{max,u}_{c,\text{EWH}}, \dots, T-1-K^\text{max,u}_{c,\text{EWH}}\right\} \label{eq:maxunblockEWH}
\end{align}

\subsubsection{EV constraints}
The \ac{DLC} with limited information considers an \ac{EV}'s charging demand as soon as a charging session starts and the customer specifies the departure settings. The constraint set includes \eqref{eq:EVhome}~--~\eqref{eq:powerEV} for all $t\in\mathcal{T}$ and \eqref{eq:SoCgoalEV1}~--~\eqref{eq:SoCminEV2} for ongoing sessions, i.e., those including time step $0$. An \ac{EV} is assumed to be disconnected ($\phi_{c,t}^\text{EV}=0$) for $t \geq t_{c,0}^\text{EV,d}$.

\subsubsection{Objective function}
To reduce the impact of prediction errors, we not only target the single highest peak in the time horizon, but also aim to flatten the overall load curve by choosing a quadratic objective function. It penalizes deviations of the total power from a specified reference value $P^\text{ref}$:
\begin{subequations}
\begin{align}
\min \;\; &\sum_{t=0}^{T-1} \left(
    P_t^\text{tot}-P^\text{ref}
\right)^2\\
\text{s.t.} \;\: \, &P_t^\text{tot} \!= P_{t}^\text{inf}
    + \!\!\sum_{c\in \mathcal{C^\text{HP}}}\!\! P^\text{HP}_{c,t} 
    + \!\!\sum_{c\in \mathcal{C^\text{EWH}}}\!\! P^\text{EWH}_{c,t}
    + \!\!\sum_{c\in \mathcal{C^\text{EV}}}\!\! P^\text{EV}_{c,t} 
    , \, \forall t \in \mathcal{T}\raisetag{1.5ex}\label{eq:lim_obj_ptot}
\end{align}
\label{eq:lim_obj}
\end{subequations}

\noindent where $P^\text{inf}_t$ are now predictions of the inflexible load at the \ac{TS} level over the optimization horizon, see Section~\ref{sec:Inflex_load_est}.

\subsection{Variants and settings}
\label{sec:formVar}
Table~\ref{tab:form} summarizes the \ac{DLC} with perfect and with limited information, and provides the parameter values for the different constraints with $\Delta t = 0.25$~h.
The values for $K^\text{block,24h}_{c,\text{HP}}$, $K^\text{max,b}_{c,\text{HP}}$, $K^\text{min,u}_{c,\text{HP}}$, and $K^\text{block,24h}_{c,\text{EWH}}$ are based on \ac{EKZ}'s standard ripple control, and the value for $K^\text{run}_{c,\text{EWH}}$ is based on the median average \ac{EWH} runtime obtained in \cite{KreftIdentifyingElectric2024}.
$\text{SoC}_c^\text{EV,min}$ is set to $0.5$ for all cases.

The implementation in the pilot corresponds to (L1) with the minimum/maximum up/down constraints formulated in the $u$-space only (cf. Section~\ref{sec:formIdealHP}), an alternative formulation for the \ac{EV} constraints \eqref{eq:SoCgoalEV1} and \eqref{eq:SoCgoalEV2}, which includes the \acs{SoC} evolution, and without \eqref{eq:minblockHP3}, allowing for potential invalid blockings at the end of the optimization horizon. This variant is referred to as (L1-Pilot), and the detailed formulation is provided in \cite{KaiserOrtsNetz2025}. The simulations evaluate the peak-shaving performance of (Ideal), (L1-Pilot), (L1), and (L2). For (L2), additional simulations are performed with different inputs for the \ac{EV} charging sessions and inflexible load profiles. More specifically, (L2-PF$_\text{EV,inf}$) assumes perfect foresight regarding \ac{EV} charging sessions and the inflexible load, (L2-PF$_\text{inf}$) assumes perfect foresight for the inflexible load only, and (L2-Shift$_{\,\text{+1h}}$) and (L2-Shift$_{\,\text{-1h}}$) use the inflexible load estimates shifted by one hour. These variants are introduced to assess the impact of future charging sessions and the quality of the inflexible load estimation.

\begingroup
\setlength{\tabcolsep}{2.5pt}
\begin{table*}
    \centering
    \caption{Overview of the available information and the constraints for the investigated cases.}
    \begin{tabular*}{\linewidth}{@{\hspace{5pt}}@{\extracolsep{\fill}}cllccc|ccccccc@{\hspace{5pt}}}
    \toprule
    & & & Ideal & \multicolumn{2}{c|}{Limited} & \multicolumn{7}{c}{Settings}\\
    & & & & L1 & L2 & $K^\text{block,24h}_{c,\cdot}$ & $K^\text{min,b}_{c,\cdot}$ & $K^\text{max,b}_{c,\cdot}$ & $K^\text{min,u}_{c,\cdot}$ & $K^\text{max,u}_{c,\text{EWH}}$ & $K^\text{run}_{c,\text{EWH}}$ & $K^\text{max,24h}_{c,\text{EV}}$\\
    \midrule
    \multicolumn{2}{l}{Objective function}  & & \eqref{eq:ideal_obj} & \eqref{eq:lim_obj} & \eqref{eq:lim_obj}\\
    \midrule
    \multirow{8}{*}{\rotatebox[origin=c]{90}{Constraints}} 
    & \ac{HP} flexibility & \eqref{eq:blockHP} -- \eqref{eq:minblockHP3} & \checkmark & \checkmark & \checkmark & 16 & 4 & 8 & 8 & & &\\
    & \ac{EWH} flexibility & \eqref{eq:blockWH} & \checkmark & \checkmark & \checkmark & 64 -- 78 & & & & & &\\
    & \ac{HP} power, actual model & \eqref{eq:tempEvoHP} -- \eqref{eq:PModHPideal} & \checkmark & & &\\
    & \ac{EWH} power, actual model & \eqref{eq:tempEvoHP} -- \eqref{eq:POnOffHPideal} for \acp{EWH} & \checkmark & & &\\
    & \ac{HP} power, estimation & \eqref{eq:powerHP} & & \checkmark & \checkmark\\
    & \ac{EWH} power, estimation L1 & \eqref{eq:blockinstHP} -- \eqref{eq:minblockHP3} for \acp{EWH}, \eqref{eq:powerEWH1} & & \checkmark & & & 19 & 56 -- 72 & 6 -- 8 & & &\\
    & \ac{EWH} power, estimation L2 & \eqref{eq:minunblockHP} -- \eqref{eq:minblockHP3} for \acp{EWH}, \eqref{eq:Krunrho1EWH} -- \eqref{eq:maxunblockEWH} & & & \checkmark & & 40 & & 12 & 40 & 6 & \\
    & \ac{EV} charging flexibility &\eqref{eq:SoCgoalEV1} -- \eqref{eq:auxvarEV}  & \checkmark & \checkmark & \checkmark & & & & & & & 3\\
    & \ac{EV} charging power & \eqref{eq:powerEV} & \checkmark & \checkmark & \checkmark \\
    \midrule
    \multirow{2}{*}{\rotatebox[origin=c]{90}{Input}} & \ac{EV} charging sessions & only ongoing (o) or all (a) & a & o & o\\
    & Inflexible load profile & estimated (e) or actual (a) & a & e & e\\
    \bottomrule
    \end{tabular*}
    \label{tab:form}
\end{table*}    
\endgroup

\section{Case study}
\label{sec:res}
In the following, we first describe the real-world pilot setup and the inflexible load, \ac{HP} power, and baseline estimation. Subsequently, we present the pilot results and provide further insights based on simulations.
\vspace{-0.4cm}
\subsection{Pilot}
\label{sec:resPil}
\subsubsection{Experimental setup and infrastructure}
\label{sec:resPilSetup}
For testing the \ac{DLC} with limited information in a real-world pilot application, load control devices were installed at $22$ single-family homes that are connected to the same \ac{TS}. Unlike the standard ripple control, which blocks groups of loads simultaneously, the new devices allow for individualized commands for each \ac{HP} and \ac{EWH}. Every $15$ minutes ($\kappa=1$), the blocking commands for the next $24$~hours ($T=96$) were computed in the central optimization using the Gurobi \cite{gurobi} solver, configured with a $5$-minute time limit and the barrier method for solving the root and node relaxations.
The resulting commands were communicated to the local devices via power line communication. A local verification module corrected the commands in case they did not satisfy the flexibility constraints \eqref{eq:blockHP} -- \eqref{eq:minblockHP3} and \eqref{eq:blockWH}, e.g., due to communication issues. The central system retrieved the actually applied blocking actions and the $15$-minute \ac{SM} measurements once a day.
Additionally, several \acp{EV} in the same municipality were virtually connected to the \ac{TS} under study, i.e., their charging profiles were added to the \ac{TS} load even though they were physically connected to a different \ac{TS}. \Ac{EV} charging was controlled via an \ac{API} that can start and stop charging and retrieve the vehicles' current status. In each iteration of the rolling horizon, the optimization considered the \acp{EV} that were available for control at the given moment. For consistency, the analyses also consider only those \ac{EV} charging sessions captured in at least one optimization window.
During the pilot period, the relevant \ac{EV} customers and all customers connected to the given \ac{TS} received a unit tariff instead of the existing two-tier tariff \cite{KaiserOrtsNetz2025}. \acp{HP} and \acp{EWH} without the new control device remained under the standard ripple control, and were considered part of the inflexible load.

\subsubsection{Inflexible load estimation and $P^\text{ref}$}
\label{sec:Inflex_load_est}
The \ac{DLC} needs a 24-hour-ahead prediction of the inflexible load $P^\text{inf}$ at the \ac{TS} level in \eqref{eq:lim_obj}. To enable such a prediction, past \ac{SM} data of every house was first disaggregated to create a database of historical flexible (\ac{HP} and \ac{EWH}) demand. Later, predicted ambient temperature and global irradiation data were used to identify the day that, out of the historical data, is most similar to the considered day. This day is referred to as \ac{MSD}. Finally, the inflexible load prediction for this day was calculated by subtracting the flexible load profiles (\ac{HP}, \ac{EWH}, and \ac{EV}, the latter through direct measurement) from the total \ac{TS} load of the \ac{MSD}.

 Preliminary tests found that using $P^\text{ref}>0$ in formulation (L1-Pilot) would result in flatter total load curves. The value of $P^\text{ref}$ in \eqref{eq:lim_obj} was set to the average total power at the \ac{TS} level for the identified \ac{MSD}.

\subsubsection{HP power estimation}
\label{sec:HP_power_estimation}
\begin{figure}
    \centering
    \includegraphics{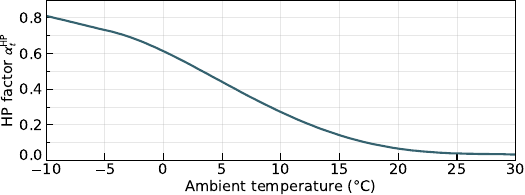}
    \caption{Estimated variation of the \acp{HP}' power demand with ambient temperature.}
    \label{fig:sim_alpha_HP}
\end{figure}

The power demand of each \ac{HP} is estimated as its nominal power multiplied by the correction parameter $ 0 \leq \alpha_t^\text{HP} \leq 1$. This parameter represents the variation of \acp{HP}' power demand with ambient temperature. It is computed from a regression model extracted from two years of historical data of the power demand of dozens of \acp{HP} in Zurich and the corresponding ambient temperature. Fig.~\ref{fig:sim_alpha_HP} displays the curve that was fitted to four-hour average temperature and power values. The regression model is used to determine $\alpha_t^\text{HP}$ given an ambient temperature forecast. This parameter is assumed to be equal for all devices.

\subsubsection{Baseline load estimation}
\label{sec:resPilBaseline}
The \ac{TS} load profile without the \ac{DLC} needs to be estimated to evaluate the \ac{DLC} performance in the pilot. This work follows the baseline load estimation approach described in \cite{KaiserPeakReduction2025}. The baseline load for a given pilot day is estimated as the actual \ac{TS} load profile, minus the \ac{EV} charging load during controlled sessions, plus the load profiles of those sessions with uncontrolled charging, minus the \ac{SM} profiles of households with an OrtsNetz load control device on the pilot day, plus the \ac{SM} profiles of those households on the corresponding \ac{MSD}. Two different \ac{MSD} matching approaches are considered to strengthen the validity of the findings. The \ac{MSD} is identified either as (a) the day with the most similar weather conditions --- as described above for the inflexible load estimation --- or (b) the day whose load profile, considering only non-participants in the pilot municipality, is most similar. The reader is referred to \cite{KaiserPeakReduction2025} for more details.

\subsubsection{Results}
\label{sec:resPilResults}
Table~\ref{tab:dates_schemes} gives the dates for evaluating injection and consumption peak reductions in summer and winter, respectively.
\begin{table}
\centering
\caption{Dates for the pilot evaluation.}
\footnotesize
\centerline{
\begin{tabular}{l | l | l l }
    \toprule
    & OrtsNetz pilot &  \multicolumn{2}{c}{Baseline period} \\
    \midrule
    Season & Dates & Scheme & Dates \\
    \midrule
    Summer
    & 29.08.24 - 04.09.24 & Ripple control  & 01.06.23 - 14.09.23 \\
    \midrule
    \multirow{2}{*}{Winter} 
    & 19.11.24 - 19.12.24 & Ripple control & 01.01.23 - 31.03.23 \\
    & 19.11.24 - 19.12.24 & No control & 20.12.23 - 27.02.24 \\
    \bottomrule
\end{tabular}
}
\label{tab:dates_schemes}
\end{table}
During the evaluated pilot periods, \SI{90.1}{\%} (resp. \SI{89.7}{\%}) of the \ac{HP} (resp. \ac{EWH}) commands used as input by the local control devices matched the signals sent by the central controller. The local verification, which was implemented to filter out invalid \ac{DLC} commands, changed the \ac{HP} and \ac{EWH} commands \SI{7.2}{\%} and \SI{16.5}{\%} of the time, respectively.
These changes stem from communication issues and a mismatch between the central and local settings for $K^\text{block,24h}_{c,\text{EWH}}$ for some \acp{EWH}, which led to additional unblockings of up to $3.5$~hours.

\begin{figure*}
    \centering
    \includegraphics[width=\linewidth]{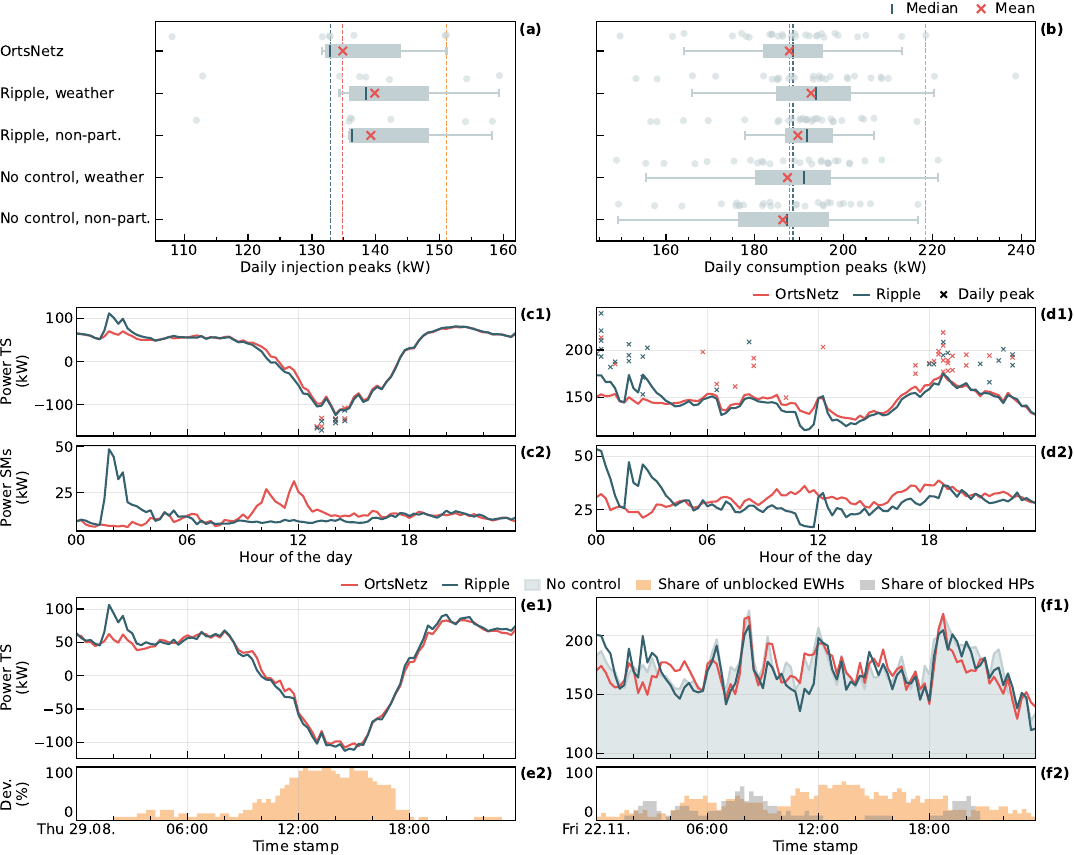}
    \caption{Pilot results for summer (left) and winter (right): Distribution of daily peaks (top), mean daily profiles and daily peak values using the \ac{MSD} based on weather data (middle), and TS load profiles and device blockings for a single day using the \ac{MSD} based on weather data (bottom).}
    \label{fig:res_pilot}
\end{figure*}
The left column of Fig.~\ref{fig:res_pilot} shows the pilot results for summer. Fig.~\ref{fig:res_pilot}(a) compares the distribution of daily injection peaks in OrtsNetz, i.e., the \ac{DLC} with formulation  (L1-Pilot), and the corresponding baseline load profiles with ripple control, according to the weather and non-participants \ac{MSD} matching approaches.
The \ac{DLC} reduced the injection peaks by $4$ or \SI{5}{kW} on average, depending on the matching approach. 
Fig.~\ref{fig:res_pilot}(c) compares the mean daily \ac{TS} power as a result of OrtsNetz with ripple control, and the corresponding aggregated profiles of \acp{SM}, which were replaced to obtain the baseline load estimate. It illustrates that the \ac{DLC} shifts the \ac{EWH} operation from nighttime to late morning. Furthermore, it suggests that greater reductions could have been achieved if the \acp{EWH} had consumed more power during the peak \ac{PV} injection hours. As shown in Fig.~\ref{fig:res_pilot}(e2), although most \acp{EWH} were unblocked from 9~am to 5~pm, they primarily demanded power in the initial hours after the unblocking signals, i.e., consumed power before the injection peaks.
Formulation (L2) addresses this deficiency and is investigated in simulations.

In winter, the baseline periods from which \acp{MSD} were selected include days when the ripple control was active and when no control scheme was in effect (see Table~\ref{tab:dates_schemes}). In the no-control setting, all devices connected to an OrtsNetz control device were unblocked for the entire day and could therefore operate freely according to their internal device controllers.
The right column of Fig.~\ref{fig:res_pilot} presents the evaluation results for winter days. \acp{HP} and other heating systems dominate the \ac{TS} demand. Hence, consumption peaks become the determining factor.
Fig.~\ref{fig:res_pilot}(b) shows that the \ac{DLC} reduced the consumption peaks by an average of $2$ to \SI{5}{kW} with respect to the ripple control. However, no peak reductions were observed compared to no control. The comparison of the \ac{SM} profiles in Fig.~\ref{fig:res_pilot}(d2) shows that the \ac{DLC} distributed the flexible demand more evenly throughout the day, with some demand concentration between 9~am and 7~pm. For a more detailed investigation, Fig.~\ref{fig:res_pilot}(f) presents the share of unblocked \acp{EWH} and blocked \acp{HP} on a single day. At the two consumption peaks of the \ac{TS}, several \acp{HP} and \acp{EWH} were blocked, yet insufficient to prevent the peaks. As reported in the load profiles comparison of Fig.~\ref{fig:res_pilot}(f1), the \ac{DLC} reduced the consumption peak experienced in the early hours, but there was no clear advantage when compared to the no-control case for other hours. During the consumption peaks, the \ac{DLC} reached almost the same demand values as those without control.

In the final project survey, none of the pilot \ac{DLC} participants reported undesired changes in controlled temperature.
This suggests that the proposed constraints provide enough room for the devices' normal operation while providing flexibility when needed. However, three survey responses indicate that future implementations should provide clear explanations of the \ac{HP} and \ac{EWH} flexibility constraints to the customers.

\subsection{Simulations}
\label{sec:resSim}
While the real-world results demonstrate technical feasibility, they may have been affected by the uncertainties in the baseline load estimate and technical issues \cite{KaiserPeakReduction2025}. The following simulations eliminate these aspects and investigate the performance of the variants introduced in Section~\ref{sec:formVar}.
\subsubsection{Setup}
\label{sec:resSimSetup}
The simulation environment emulates the \ac{TS} used to test the \ac{DLC} in the real-world pilot.
The inflexible load profile $P^{\text{inf}}$ is derived by (i)~subtracting the \ac{SM} profiles of controllable households from the total \ac{TS} load, (ii)~adding the average \ac{SM} profile of inflexible customers without electric heating, \ac{EWH}, \ac{EV}, and \ac{PV}, scaled by the number of controlled households, and (iii)~subtracting the generation profile of two separately measured \ac{PV} installations, scaled to match the controlled households' \ac{PV} generation.
The parameters for the \ac{HP} and \ac{EWH} models (see Section~\ref{sec:formIdeal}) are derived as described in \cite{KaiserDynamicGrid2024}, except that (i)~the \ac{EWH} storage tank volumes are taken from installation data, and (ii)~the \ac{HP} storage tanks are sized to be able to bridge between $1$ and $2$ hours of blocking at the maximum heat demand, similar to \cite{HeiderGridReinforcement2023}. \acp{HP} with a nominal electrical power greater than or equal to $7$~kW are modeled as modulating \acp{HP}, with $P_{c,\text{min}}^\text{HP} = 0.3 \cdot P_{c,\text{nom}}^\text{HP}$. All other \acp{HP} are assumed to be on-off-controlled.
The \ac{EV} data and charging characteristics are based on real-world data from the pilot with the additional assumptions that each \ac{EV} has a fixed charging power and battery capacity, and that the charging efficiency is \SI{90}{\%}. The recorded \ac{EV} plug-in times, \acs{SoC} at plug-in, and \acs{SoC} at departure are used to determine each session's requirements.

Simulations are carried out for two periods: one week in summer (August 29th to September 4th, 2024) and one in winter (December 9th to December 15th, 2024). Devices are initialized by running the simulation without external control for two additional days prior to the start of each period.

\begin{figure}
    \centering
    \includegraphics{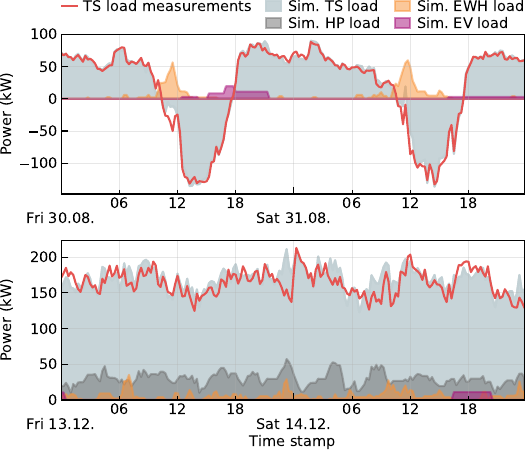}
    \caption{Measured and simulated load for two summer days (top) and two winter days (bottom).}
    \label{fig:sim_res_actual_vs_sim_dlc}
\end{figure}

As visualized in Fig.~\ref{fig:sim_res_actual_vs_sim_dlc}, the simulated \ac{TS} load is generally aligned with the real-world measurements. In this comparison, \acp{HP} and \acp{EWH} were simulated using the blocking commands that were applied during the pilot, and the \ac{EV} load represents immediate charging.

The computation time limit to solve the optimization problem in the perfect knowledge case is set to $4$~hours, and the optimization window is shifted by $6$~hours between iterations ($\kappa=24$). For all other cases, the time limit remains the same as in the pilot, i.e., $5$~minutes, and the optimization window is shifted by $1$~hour ($\kappa=4$). The solution of the previous iteration serves as a warm start for the next iteration, except for (L$1$-Pilot), which --- consistent with the pilot --- does not use a warm start.
The estimations for $P^\text{inf}$ and $P^\text{ref}$ are the same as during the pilot for (L1-Pilot), (L1), and (L2). For the variants (L2-PF$_\text{EV,inf}$) and (L2-PF$_\text{inf}$), $P^\text{inf}$ is equal to the actual inflexible load used in the simulations, and $P^\text{ref}$ is fixed at $0$\,\si{kW} and \SI{160}{kW} in summer and winter, respectively.

The optimization results may vary based on the solver's solution path if there are multiple optimal solutions or when the computation time limit is reached prior to obtaining the optimal solution. To account for this variability, each variant is evaluated across ten runs with distinct random number seeds.

\subsubsection{Results}
\label{sec:resSimResults}
\begin{figure}
    \centering
    \includegraphics{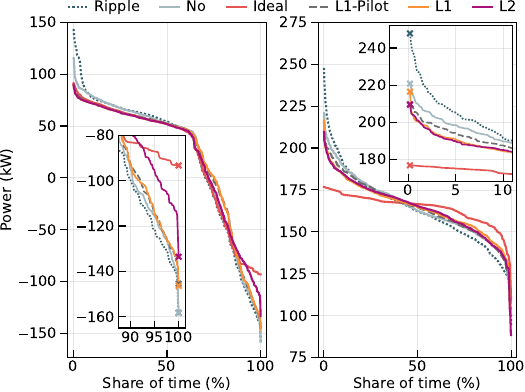}
    \caption{Duration curves for the simulations in summer (left) and winter (right).}
    \label{fig:sim_res_dur}
\end{figure}

\begingroup
\renewcommand{\arraystretch}{1.2}
\begin{table}
\centering
\caption{
Simulation results for injection and consumption peaks in summer and winter, respectively.}
\footnotesize
\settowidth{\mylongcol}{$\overline{\Delta P^{\text{max}}}/P^{\text{EWH}}$}
\begin{tabular*}{\linewidth}{@{\hspace{3pt}}@{\extracolsep{\fill}}
l@{\hspace{0.2cm}}l
r@{\hspace{\mysep}}r
w{r}{\dimeval{\mylongcol/3*2-\mysep}}@{\hspace{\mysep}}w{l}{\dimeval{\mylongcol/3-\mysep}}
r@{\hspace{\mysep}}r@{\hspace{3pt}}}
\toprule
& Scheme & \multicolumn{2}{c}{$\overline{\Delta P^{\text{max}}}$} & \multicolumn{2}{c}{$\overline{\Delta P^{\text{max}}}/P^{\text{EWH}}$} & \multicolumn{2}{c}{$\Delta P^{\text{max}}$} \\
& & \multicolumn{2}{c}{(kW)} & \multicolumn{2}{c}{(\%)} & \multicolumn{2}{c}{(kW)} \\
\midrule
\multirow{5}{*}{\rotatebox[origin=c]{90}{Summer}} 
& Ripple & $-2.96$ &  & $-3.2$ &  & $0.00$ &  \\
& L1-Pilot & $5.96$ &  & $6.4$ &  & $9.25$ &  \\
& L1 & $6.31$ &  & $6.8$ &  & $11.85$ &  \\
& L2 & $23.73$ &  & $25.4$ &  & $24.85$ &  \\
& Ideal & $54.81$ &  & $58.8$ &  & $64.60$ &  \\
\midrule
\multirow{5}{*}{\rotatebox[origin=c]{90}{Winter}}
& Ripple & $-10.81$ &  & \text{-} &  & $-27.05$ &  \\
& L1-Pilot & $6.69$ & ($5.71$) & \text{-} &  & $-0.46$ &  \\
& L1 & $11.68$ & ($10.41$) & \text{-} &  & $9.65$ & ($8.96$) \\
& L2 & $11.51$ & ($10.06$) & \text{-} &  & $13.00$ & ($11.41$) \\
& Ideal & $41.37$ & ($35.82$) & \text{-} &  & $44.10$ & ($38.64$) \\
\bottomrule
\end{tabular*}
\label{tab:sim_sum}
\end{table}
\endgroup
\begingroup
\renewcommand{\arraystretch}{1.2}
\begin{table}
\centering
\caption{
Simulation results with different inputs for the EV charging sessions and inflexible load profiles.}
\footnotesize
\settowidth{\mylongcol}{$\overline{\Delta P^{\text{max}}}/P^{\text{EWH}}$}
\begin{tabular*}{\linewidth}{@{\hspace{3pt}}@{\extracolsep{\fill}}
l@{\hspace{0.2cm}}l
r@{\hspace{\mysep}}r
w{r}{\dimeval{\mylongcol/3*2-\mysep}}@{\hspace{\mysep}}w{l}{\dimeval{\mylongcol/3-\mysep}}
r@{\hspace{\mysep}}r@{\hspace{3pt}}}
\toprule
& Scheme & \multicolumn{2}{c}{$\overline{\Delta P^{\text{max}}}$} & \multicolumn{2}{c}{$\overline{\Delta P^{\text{max}}}/P^{\text{EWH}}$} & \multicolumn{2}{c}{$\Delta P^{\text{max}}$} \\
& & \multicolumn{2}{c}{(kW)} & \multicolumn{2}{c}{(\%)} & \multicolumn{2}{c}{(kW)} \\
\midrule
\multirow{5}{*}{\rotatebox[origin=c]{90}{Summer}} 
& L2-Shift$_{\,\text{+1h}}$ & $15.72$ &  & $16.9$ &  & $13.41$ &  \\
& L2-Shift$_{\,\text{-1h}}$ & $15.18$ &  & $16.3$ &  & $20.78$ &  \\
& L2 & $23.73$ &  & $25.4$ &  & $24.85$ &  \\
& L2-PF$_\text{inf}$ & $25.96$ &  & $27.8$ &  & $31.97$ &  \\
& L2-PF$_\text{EV,inf}$ & $25.08$ &  & $26.9$ &  & $27.96$ &  \\
\midrule
\multirow{5}{*}{\rotatebox[origin=c]{90}{Winter}}
& L2-Shift$_{\,\text{+1h}}$ & $6.84$ & ($5.08$) & \text{-} &  & $-0.72$ & ($-1.92$) \\
& L2-Shift$_{\,\text{-1h}}$ & $5.60$ & ($4.38$) & \text{-} &  & $3.45$ & ($-3.84$) \\
& L2 & $11.51$ & ($10.06$) & \text{-} &  & $13.00$ & ($11.41$) \\
& L2-PF$_\text{inf}$ & $18.93$ & ($16.83$) & \text{-} &  & $12.60$ &  \\
& L2-PF$_\text{EV,inf}$ & $18.46$ & ($15.77$) & \text{-} &  & $13.55$ & ($13.08$) \\
\bottomrule
\end{tabular*}
\label{tab:sim_sum_inflx}
\end{table}
\endgroup
\begin{figure}
    \centering
    \includegraphics{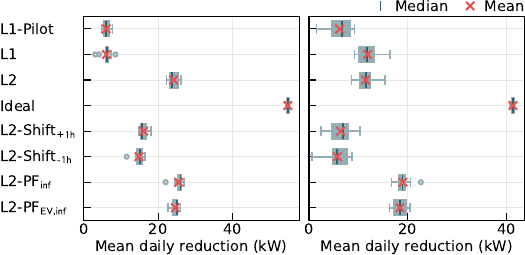}
    \caption{Distribution of the mean daily injection peak reductions in summer (left) and consumption peak reductions in winter (right) across ten different seeds.}
    \label{fig:sim_diff_seeds}
\end{figure}
Tables~\ref{tab:sim_sum} and \ref{tab:sim_sum_inflx} summarize the peak reduction results for the different formulations and for different inputs to (L2), respectively. The metrics are computed with respect to the simulation without any external control. $\overline{\Delta P^{\text{max}}}$ is the average of the peak reductions on the evaluated days, while $\Delta P^{\text{max}}$ is the maximum peak reduction, i.e., the difference between the highest values across the entire week with the given scheme and without control. Since in summer, \acp{EWH} are the main source of flexibility in the given setup, we additionally present the results for the ratio $\overline{\Delta P^{\text{max}}}/P^{\text{EWH}}$, where $P^{\text{EWH}}$ is the sum of the controlled \acp{EWH}' nominal power values. This allows to generalize the results to other system configurations. Values in parentheses show the results with immediate \ac{EV} charging, i.e., without using the \acp{EV}' flexibility. These values are shown if they differ from the results with \ac{EV} flexibility. All values represent the median of the results across the ten runs with different seeds. 
Additionally, Fig.~\ref{fig:sim_diff_seeds} presents the variability of $\overline{\Delta P^{\text{max}}}$ across the different evaluation runs. Although variation exists within each variant, the relative shifts among the distributions are consistent with those of the median values, confirming that the medians serve as a representative metric for subsequent analysis. Finally, Fig.~\ref{fig:sim_res_dur} shows load duration curves for the different formulations for summer and winter. Each curve displays the average power in every $15$-minute time step of the evaluation period in descending order. For each formulation, the run closest to the median value for $\overline{\Delta P^{\text{max}}}$ is displayed, selecting the run with the higher mean reduction if the median falls between two runs.

These simulations yield five main observations. First, the results suggest that removing the ripple control may reduce peaks, which is in line with the real-world results (cf.~Fig.~\ref{fig:res_pilot}(b)) and the results in \cite{KaiserPeakReduction2025}. Without ripple control, the \ac{EWH} load is spread out throughout the day, leading to slight injection peak reductions in summer, and more significant demand peak reductions in winter.

\begin{figure}
    \centering
    \includegraphics{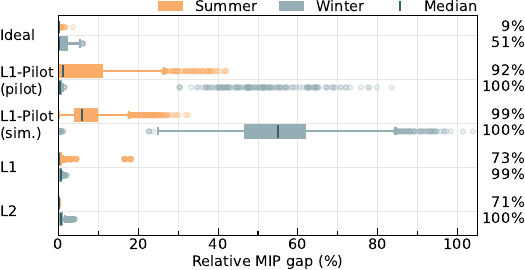}
    \caption{Relative MIP gap per iteration during the simulated periods. The percentages on the right report how often the time limit was reached before achieving a gap below the default value of \SI{0.01}{\%}.}
    \label{fig:sim_res_mip}
\end{figure}

Second, (L1-Pilot) performs similarly to (L1) in summer, and slightly worse than (L1) in winter. Fig.~\ref{fig:sim_res_mip} reveals that high \ac{MIP} gaps occurred for (L1-Pilot), both in simulations and during the pilot. In addition to the differences described in Section~\ref{sec:formVar}, the implementation for (L1-Pilot) did not use auxiliary variables $P_t^\text{tot}$ in \eqref{eq:lim_obj}. Instead, it included the products in \eqref{eq:powerEV}, \eqref{eq:powerHP}, and \eqref{eq:powerEWH1} directly in the objective function. High gaps occurred particularly when the solver linearized the resulting products of binary variables (the parameter \texttt{PreQLinearize} was at its default setting, i.e., automatic choice). Introducing the variables $P_t^\text{tot}$, $v_{c,t}^\cdot$, $w_{c,t}^\cdot$, and warm-starting the optimization reduces the gaps in the other cases.

\begin{figure*}
    \centering
    \includegraphics{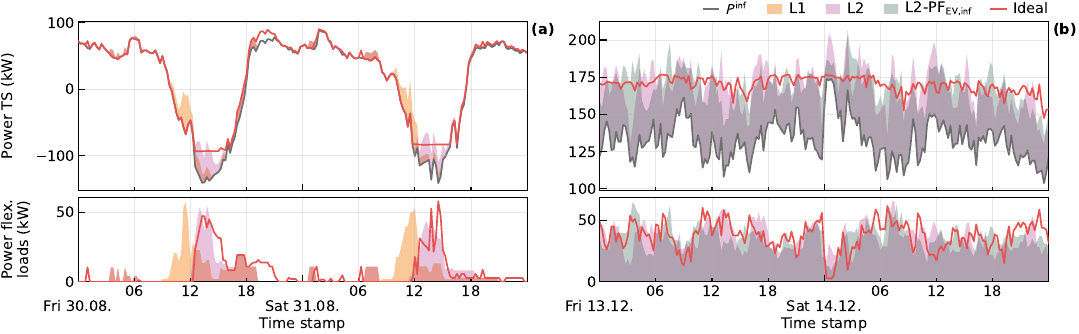}
    \caption{Simulated \ac{TS} load profiles for a selection of cases over two summer days (a) and two winter days (b). For each variant, the run closest to the median value for $\overline{\Delta P^{\text{max}}}$ is displayed.}
    \label{fig:sim_res_time_series}  
\end{figure*}

Third, formulation (L2) leads to higher injection peak reductions than (L1) due to the better representation of the \ac{EWH} operation when unblocked. Formulation (L1) does not include an upper limit on the unblocking duration of \acp{EWH}. Furthermore, it assumes that each \ac{EWH} operates at rated power whenever it is unblocked. Thus, the \ac{DLC} initiates \ac{EWH} unblocking as soon as the estimated inflexible load drops below the specified reference value $P^\text{ref}$. In contrast, (L2) accounts for the fact that \acp{EWH} primarily demand power at the beginning of an unblocked period. As illustrated in Fig.~\ref{fig:sim_res_time_series}(a), explicitly modeling this behavior helps to concentrate \ac{EWH} demand at the hours of maximum \ac{PV} injection.

Fourth, Table~\ref{tab:sim_sum_inflx} shows that the quality of the inflexible load estimates affects performance. The average peak reduction $\overline{\Delta P^{\text{max}}}$ decreases when using the inflexible load estimates shifted by one hour as input, while it increases when using the actual inflexible load as input. Perfect knowledge regarding future \ac{EV} charging sessions does not lead to further improvements in the given setup.

Finally, the recommended formulation with limited information, (L2), could realize \SI{43.3}{\%} and \SI{27.8}{\%} of the potential average daily peak reduction $\overline{\Delta P^{\text{max}}}$ with (Ideal) in summer and winter, respectively.


\section{Discussion}
\label{sec:discussion}
The simulation results demonstrate that the proposed \ac{DLC} schemes with limited information reduce injection and consumption peaks at a \ac{TS} level. Furthermore, they show that the improved formulation (L2) yields greater injection peak reductions than (L1), due to a more accurate representation of the \ac{EWH} operation when unblocked. Additionally, the real-world results demonstrate that the approach is technically feasible and preserves customer comfort. However, there are some limitations that are discussed in the following and may be addressed in future work. We focus on the formulation with limited information, which is the most relevant for practical application.

First, the results for different inflexible load inputs highlight that the \ac{DLC} performance correlates with the quality of the inflexible load estimates. The comparison between (L2) and (L2-PF$_\text{inf}$) suggests that improving the proposed inflexible load estimation approach could enhance performance. If available, this can be achieved by leveraging device-level demand measurements rather than disaggregation to determine the historical flexible demand, which is subtracted from the total \ac{TS} load. Furthermore, alternative disaggregation and forecasting methods may be investigated.

\begin{figure}
    \centering
    \includegraphics{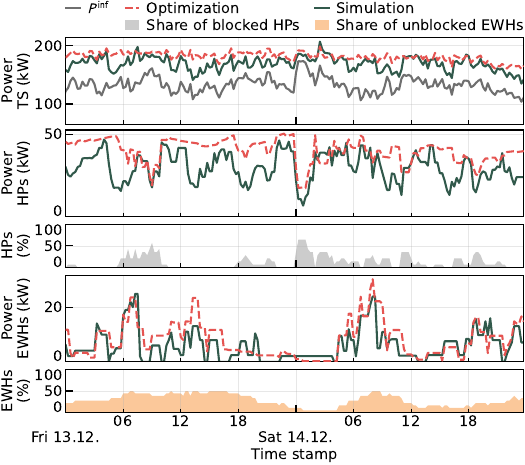}
    \caption{Comparison between the optimization and simulation results for the formulation with perfect foresight, \mbox{(L2-PF$_\text{EV,inf}$)}.}
    \label{fig:sim_res_opt_vs_sim}   
\end{figure}

Second, the \ac{HP} power estimation may be improved. A comparison between the winter results for (L2-PF$_\text{EV,inf}$) and (Ideal) reveals a significant performance gap, even under perfect knowledge of the inflexible load and future \ac{EV} charging sessions. To investigate this mismatch, Fig.~\ref{fig:sim_res_opt_vs_sim} compares the optimization and simulation results for the formulation (L2-PF$_\text{EV,inf}$) on two winter days, which are also visualized in Fig.~\ref{fig:sim_res_time_series}(b). The optimization and simulation results deviate because the optimization model relies on simplified \ac{EWH} and \ac{HP} power estimations (cf. Section~\ref{sec:formLimited}), whereas the simulation environment, to which the resulting switching commands are applied, uses the more detailed models that were introduced as part of the ideal formulation (cf. Section~\ref{sec:formIdeal}). The figure shows that the \ac{EWH} power demand in the simulations aligns with the estimates in the optimization, whereas there are significant differences for the \ac{HP} power. Future work could refine the \ac{HP} power estimation and model power rebounds after blocking periods~\cite{Mueller2019}.

Third, the flexible \ac{EV} charging load is low in the given case study (on average $10$\,\si{kWh} and \SI{20.5}{kWh} per day in the summer and winter simulations, respectively). However, additional \acp{EV} can easily be integrated into the optimization model. It is expected that a larger fleet of controllable \acp{EV} will make the benefits of a \ac{DLC} even more evident. The \ac{DLC} can shift the flexible \ac{EV} demand to times with a low inflexible load, thereby flattening the evening peak typically observed with uncontrolled residential charging \cite{HerreraModelingCharging2025}. At the same time, forecasting future sessions may become necessary for larger \ac{EV} fleets to account not only for ongoing sessions, but also for future arrivals.

Finally, further investigations showed that the quadratic objective function \eqref{eq:lim_obj} may yield flexible load reductions at off-peak times when the total flexible energy in the optimization horizon is not fixed. This applies to the \ac{HP} power estimation, which is a function of the temperature-dependent \ac{HP} parameter $\alpha_t^\text{HP}$. In this case, the objective function value may be smaller when reducing overall consumption rather than reducing the flexible load at times when the inflexible load is high. The choice of $P^\text{ref}$ impacts this trade-off. Furthermore, note that even if the total energy is fixed, the quadratic objective function may not yield the smallest peak value (consider, e.g., three time steps with total power $(6, 3, 2)$ compared to $(5, 5, 1)$, with $P^\text{ref} = 0$).

In addition to these refinements, the proposed model can be generalized to additional device types. For example, \ac{AC} loads can be integrated in a similar manner to \acp{HP}. Their integration merely requires adjusting the parameter settings for the minimum and maximum blocking durations and updating the regression model to estimate the temperature-dependent power estimation factor $\alpha_t$. Similarly, electric storage heaters can be modeled like \acp{EWH}, substituting the fixed operating duration $K^\text{run}_{c}$ with a temperature-dependent duration. Finally, deferrable loads, such as dishwashers or washing machines, can be integrated by using customer-defined completion deadlines, similar to the departure time for \acp{EV}. Combined with a given or estimated uninterruptible operating duration, similar to $K^\text{run}_{c}$ for \acp{EWH}, the \ac{DLC} then determines the optimal starting times.

\section{Conclusion}
\label{sec:conc}

The proposed \acp{DLC} with limited information (L1 and L2) optimize the unblocking schedule of available flexible loads to flatten the 24-hour total load curve. These formulations use a prediction of the inflexible load, obtained from historical data, and the flexibility constraints of \acp{HP}, \acp{EWH}, and \acp{EV}. A real-world pilot implementation demonstrated both the technical feasibility of the proposed \ac{DLC} and the preservation of customers' comfort, confirming that the approach enables utilities to leverage the flexibility of major loads for peak reduction with only a few changes to the existing infrastructure. Furthermore, simulation results showed that formulation (L2) addresses the deficiency of (L1) in modeling the \ac{EWH} power demand. This improved formulation with limited information could achieve \SI{43.3}{\%} and \SI{27.8}{\%} of the potential average daily peak reduction with perfect knowledge in summer and winter, respectively.
Future research could refine the proposed \ac{DLC} by improving the inflexible load and \ac{HP} power estimations, or extend the formulation to incorporate grid constraints and additional device types. Moreover, future work could evaluate the performance across diverse system configurations, such as varying numbers of flexible devices or different underlying inflexible load profiles.

\section*{Acknowledgments}
We thank Marina Gonz\'alez Vay\'a and Ludger Leenders from EKZ for enabling the real-world implementation, and Markus Kreft and Thorsten Staake from ETH Zurich for valuable discussions and their support. We also gratefully acknowledge the contributions of Matteo Guscetti, Hendrik Lohse, Federico Lo Curto, and everyone involved in the real-world implementation.

\bibliographystyle{IEEEtran}
\bibliography{refs}
\vfill
\end{document}